\documentclass[aps,prl,superscriptaddress,showpacs,twocolumn]{revtex4}
\usepackage{epsfig}
\begin{document}
\title{Theory of charge sensing in quantum-dot structures}
\author{Richard Berkovits}
\affiliation{The Minerva Center, Department of Physics,
    Bar-Ilan University, Ramat-Gan 52900, Israel}
\author{Felix von Oppen}
\affiliation{Institut f\"ur Theoretische Physik, Freie Universit\"at Berlin, Arnimallee
14, 14195 Berlin, Germany}
\author{Yuval Gefen}
\affiliation{Department of Condensed Matter Physics, The Weizmann
Institute of Science, Rehovot 76100, Israel}
\date{November 15, 2004, version 4.0}

\begin{abstract}

Charge sensing in quantum-dot structures is studied
by an exactly solvable reduced model and numerical density-matrix 
renormalization group methods. Charge sensing is characterized by the 
repeated cycling of the occupation of current-carrying states due to the capacitive 
coupling to trap states which are weakly coupled to the leads. In agreement with 
recent experiments, it results in a variety of 
characteristic behaviors ranging from asymmetric Coulomb-blockade peaks to sawtooth- and 
dome-like structures. 
Temperature introduces distinct asymmetric smearing of these features and 
correlations in the conductance provide a fingerprint of charge-sensing behavior.
\end{abstract}

\maketitle

{\it Introduction.}---Within the orthodox picture of the Coulomb blockade
regime, subsequent Coulomb blockade
peaks are due to the filling of consecutive single-particle states
\cite{alhassid00}. Once a state is filled, it remains so.
Although this picture successfully
describes various transport properties of weakly-coupled
meso- and nanoscopic systems, there has been much interest in identifying
situations in which the orthodox picture fails and a
``dynamical'' behavior of the occupations of single-particle orbitals emerges.

One of the earliest pertinent examples was pointed out by 
Kuznetsov {\it et al.\ }\cite{kuznetsov96} who considered the
filling of localized states
in a barrier. They have shown that as the gate 
voltage increases, a localized
state may first fill and then vacate, once a different localized state
is occupied. This behavior is manifested in the conduction through
the barrier by the reappearance of the {\it same} conduction peak.
A new wave of interest
was motivated by the correlations observed in the transmission phase
through a quantum dot \cite{yacoby95}. An attractive explanation for these 
correlations is that a number of successive transmission peaks through the dot is 
carried by the same state \cite{oreg97,hackenbroich97,silvestrov00}. 
This, of course, requires the population of this
state to be repeatedly cycled. Various
mechanisms, which lie beyond the orthodox picture, have been proposed 
\cite{oreg97,hackenbroich97,silvestrov00}.

In this paper, we show that repeated filling of a single-particle state
is in fact a rather generic phenomenon. We find that it occurs whenever there
exist traps in the system, either by accident or by specific design of a quantum-dot
structure. This phenomenon has been seen by Lindemann {\it et al.} \cite{lindemann02}
and in tailored structures by Johnson {\it et al.} \cite{johnson03} and Kobayashi {\it et al.}
\cite{kobayashi04}. 
Consistent with these experiments, we observe that repeated
filling of a given
single-particle state as function of gate voltage can be reflected in the 
conductance in many different ways, ranging from essentially no signature to sawtooth- 
or dome-like structures to asymmetric Coulomb-blockade peaks. 
The underlying mechanism termed {\it charge sensing} in Ref.\ \cite{johnson03}, is based
on the capacitive coupling between the traps and the conducting channel.

We first obtain our results within a reduced, exactly solvable model which captures the 
essential features of the phenomenon. As a by-product 
we show how the mechanism previously proposed by 
Silvestrov and Imry \cite{silvestrov00} is a limiting case of our more general approach. 
Relying on numerical results obtained by a density-matrix renormalization-group (DMRG)
method \cite{white93,berkovits04}, 
we subsequently discuss how our results are modified when relaxing various 
restrictions of the exactly solvable model.

\begin{figure}\centering
\epsfxsize7cm\epsfbox{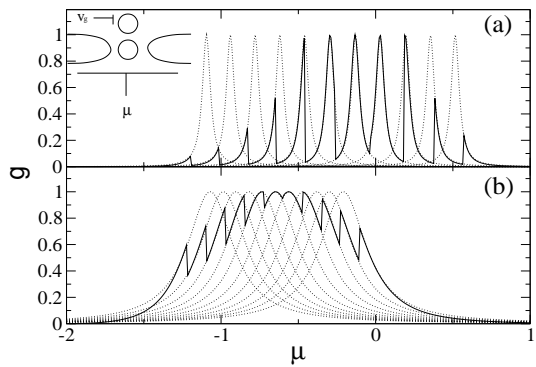} \vskip .3truecm \caption{
Conductance (full line) through a single connected state with ten 
disconnected states as function of chemical potential. 
(a) Intermediate-coupling case: $\epsilon=-1.08$, $t=0.2$, $U=0.16$
and $e_{m}=-1.2+0.02m$
($m=1, \ldots, 10$). (b) Strongly-connected case:
$\epsilon=-1$, $t=0.5$, $U=0.08$, $e_{m}=-1.24+0.04m$.
Dotted curves: Conductances for fixed
occupations $n_d=0, \ldots, 10$ of the disconnected states.
As $n_d$ changes with $\mu$, the conductance (full line)
switches accordingly. Deviations in the tails stem from the finite 
band width. Inset: Schematic of the connected 
and disconnected dots.
} \label{fig1}
\end{figure}

{\it Model.}---Motivated by the experiment of Ref.\ \cite{johnson03},
we consider a reduced model of the charge-sensing setup as shown in Fig.\ \ref{fig1}.
A quantum dot is coupled to two leads. This {\it connected} quantum dot 
is coupled electrostatically to a {\it disconnected} dot in its vicinity.
Within our reduced model, we make the following assumptions: (i) There is no tunneling between 
the disconnected dot and the leads or the connected dot. (ii) Transport through the connected dot
is carried by a single state. (iii) The disconnected dot may have many single-particle levels.
We emphasize that this model may also represent a quantum point contact with nearby traps, 
as well as transport through a single quantum dot in which one state is much more strongly 
coupled to the leads than the others.

Our reduced model is defined by the Hamiltonian
\begin{equation}
   H = H_{\rm dots} + H_{\rm leads} + H_{\rm mix}.
   \label{hamiltonian}
\end{equation}
Here, the Hamiltonian $H_{\rm dots}$ of the dots involves the level energy
$\epsilon$ of the connected dot (with creation operator
$\alpha^\dagger$) and $e_m$ of the disconnected dot (with creation operator
$d^\dagger_m$) together with the charging energies $U_d$ and $U_m$ 
for the disconnected dot and for the mutual capacitive coupling of both dots.
For simplicity, we will assume these charging energies to be equal. For spinless
electrons this gives 
$H_{\rm dots} = \epsilon \alpha^\dagger\alpha 
+\sum_{m=1}^{N_d} e_m d_{m}^\dagger d_m + U[ n_d(n_d-1)/2 + n_{\alpha} n_d]$,
where $n_{\alpha} = \alpha^\dagger\alpha$ and
$n_d = \sum_{m=1}^{N_d} d_{m}^\dagger d_m$.
The lead Hamiltonian is $H_{\rm leads} = \sum_{k,\lambda=L,R} E_{k\lambda} c_{k\lambda}^{\dagger}
c_{k\lambda}$ and the tunneling between lead and connected dot is described by
$H_{\rm mix} = \sum_{k\lambda} t_{k\lambda} \alpha^\dagger c_{k\lambda} + {\rm h.c.}$

Generally, Eq.\ (\ref{hamiltonian}) is a many-particle Hamiltonian.
Note, however, that there is no charging term for the connected dot since it can be only
singly occupied. When combined with the assumption of no tunneling to and from the 
disconnected dot, this fact makes our reduced model {\it exactly solvable}.
Since the occupations $\langle d_m^\dagger d_m \rangle$ of the disconnected-dot states can 
only take on the values $0$ or $1$, we can treat these operators 
as c-numbers in the Hamiltonian Eq.\ (\ref{hamiltonian}). Thus,
the Hamiltonian can be regarded as a set of $2^{N_d}$ {\it single-particle}
Hamiltonians, one for each possible set $\{n_m \}$ of occupation numbers 
$d_m^\dagger d_m$,
\begin{eqnarray}
H_{\{n_m \}} &=& (\epsilon + U n_d) \alpha^\dagger \alpha +
\sum_m e_{m} n_{m} 
\nonumber \\
&+& U n_d(n_d-1)/2  + H_{\rm leads} + H_{\rm mix} .
\label{hamiltonian1}
\end{eqnarray}
We first treat the zero temperature limit.
The corresponding thermodynamic potentials
at a given chemical potential $\mu$ are
\begin{eqnarray}
\Omega_{\{n_m \}} = H_{\{n_m \}}
-\mu(\alpha^\dagger \alpha + n_d + \sum_{k,\lambda=L,R} c_{k\lambda}^{\dag}
c_{k\lambda}).
\label{free}
\end{eqnarray}
The ground state occupation $\{n_m \}$ of the dots can now be found by determining  the 
configuration with the lowest thermodynamic potential. Since $\Omega_{\{n_m \}}$  is a single-particle thermodynamic potential, this can be done by calculating $\Omega_{\{n_m \}}$ for all $2^{N_d}$ 
possible configurations.

{\it Analytical treatment.}---Focusing on the essential physics, we present a full analytic treatment
of the thermodynamic potentials in Eq.\ (\ref{free}) for a {\it single} disconnected
state of energy $e_1$. The extension to several disconnected states is straightforward. 
For a single disconnected state, there are two different possibilities 
$n_d=0$ and $n_d=1$ with corresponding thermodynamic potentials $\Omega_0$ and $\Omega_1$.
As the chemical potential (gate voltage) $\mu$ increases, the disconnected state
will eventually be filled at $\mu=\mu_{\rm switch}$ when the condition $\Omega_0=\Omega_1$ is fulfilled. 
This switch in the occupation of the disconnected state is accompanied by an 
abrupt change in the occupation of -- and hence the conductance through -- 
the connected dot. It is this general phenomenon that 
is referred to as {\it charge sensing}. We now proceed with a quantitative 
analysis.

The Hamiltonians $H_0$ and $H_1$, associated with empty and occupied disconnected state, 
are single-particle Hamiltonians with eigenstates $|\psi_j^0\rangle$ and $|\psi_j^1\rangle$
and eigenenergies $\{\epsilon^0_j\}$ and $\{\epsilon^1_j\}$. Thus, we can define 
the density of states on the connected dot 
$\nu^{0 (1)}(\varepsilon)= \sum_j \langle\psi^{0 (1)}_j| \alpha^\dag \alpha|
\psi^{0 (1)}_j \rangle \delta(\varepsilon-\epsilon^{0 (1)}_j)$,
and on the leads ${\cal N}^{0 (1)}(\varepsilon)=
\sum_j \langle\psi^{0 (1)}_j| \sum_{k\lambda} c_{k\lambda}^{\dag} c_{k\lambda}|
\psi^{0 (1)}_j \rangle \delta(\varepsilon-\epsilon^{0 (1)}_j)$.  
We now express the relevant thermodynamic potentials as

\begin{eqnarray}
\Omega_{n_d} = (e_1 - \mu)\delta_{n_d,1} + \int_{-\infty}^{\mu}\! d\varepsilon (\varepsilon-\mu)
 (\nu^{n_d}(\varepsilon)+{\cal N}^{n_d}(\varepsilon)) .
\nonumber
\end{eqnarray}
Assuming that the density of states in the lead varies only on scales large compared
to both $U$ and the width $\Gamma$ of the connected state, 
we can write $\nu^0(\varepsilon)= (\Gamma/2 \pi)
((\varepsilon-\epsilon)^2+(\Gamma/2)^2)^{-1}$ and
$\nu^1(\varepsilon)= \nu^0(\varepsilon-U)$. Moreover, a finite number of disconnected states
will leave the continuum of lead states essentially unaffected so that 
${\cal N}^0 = {\cal N}^1$. 
Thus, we have
\begin{equation}
\Omega_{1} -\Omega_{0}= e_1 -\mu +
\int_{-\infty}^{\mu}d\varepsilon (\varepsilon-\mu)
(\nu^0(\varepsilon-U)-\nu^0(\varepsilon))
\label{free2}
\end{equation}
and obtain 
\begin{eqnarray}
\Omega_{1} -\Omega_{0} &=&
e_1 -\mu + U/2 + {{\mu-\epsilon}\over{\pi}}
\arctan \left({{2(\mu-\epsilon)}\over{\Gamma}}\right)
\nonumber\\
&-& {{\mu-U-\epsilon}\over{\pi}}
\arctan \left({{2(\mu-U-\epsilon)}\over{\Gamma}}\right)
\nonumber \\
&+& {{\Gamma}\over{4 \pi}} \ln
\left({{(\mu-U-\epsilon)^2+(\Gamma/2)^2}\over
{(\mu-\epsilon)^2+(\Gamma/2)^2}}\right)
\label{free3}
\end{eqnarray}
upon performing the integration.

We first consider the limit of a {\it weakly-connected} dot for which the distances of 
the connected dot energies from the chemical potential $\epsilon-\mu$ (for $n_d=0$),  
and $\epsilon+U-\mu$ (for $n_d=1$) are large 
compared to the level width $\Gamma$. Then, Eq.\ (\ref{free3}) simplifies to
$\Omega_{1}-\Omega_{0}\simeq e_1-\epsilon +(\Gamma/2\pi)\ln(|\mu-
\epsilon-U|/|\mu-\epsilon|)$, in agreement with the many-body perturbation 
theory result of Silvestrov and Imry \cite{silvestrov00}. This implies that the switching
occurs on the Coulomb-blockade plateau ($\langle n_{\alpha} \rangle$ is an integer);
hence there is no effect on the conductance at 
$\mu_{\rm switch}$. However, the switch leads to repeated appearances of the same 
Coulomb-blockade peak \cite{silvestrov00}.

We now turn to the situation when the broadening $\Gamma$ is comparable ({\it intermediate} coupling)
or larger ({\it strongly-connected}) than the distances of the dot energies from the chemical potential 
$\mu$. In these cases, the perturbation 
theory of Ref.\ \cite{silvestrov00} fails, while our general solution Eq.\ (\ref{free3}) still applies.
Specifically as $\mu$ sweeps across $\mu_{\rm switch}$ from below, the occupation of the connected-dot state 
decreases abruptly from 
$n_{\alpha}=1/2+\arctan[2 (\mu_{\rm switch}-\epsilon)/\Gamma]/\pi$ to $n_{\alpha}=1/2+\arctan[2 (\mu_{\rm switch}-\epsilon-U)/\Gamma]/\pi$. In the limit of a strongly-connected state, Eq.\ (\ref{free3})
leads to the explicit solution $\mu_{\rm switch}\simeq e_1 + U/2$.

Instead of the changes in the occupations $n_{\alpha}$, we focus directly on the experimentally more accessible
(dimensionless) conductances $g$ of the quantum-dot structure. For the specific case of a connected quantum 
dot with two symmetrically-coupled single-channel leads, the Friedel sum rule can be exploited in the usual way
to derive the relation $g=\sin^2(\pi n_{\alpha})$ \cite{datta97}. Thus, the jump in the occupation $n_{\alpha}$ at 
$\mu_{\rm switch}$ translates directly into a jump in the conductance (unless $\Delta n_{\alpha} =1$, which happens 
for a weakly-connected dot). 

{\it Numerical results}.---Representative traces of the conductance as a function of $\mu$ are 
shown in Fig.\ \ref{fig1}. These plots are based on a generalization of the above analytic results 
to the case of an arbitrary number $N_d$ of disconnected states. The most striking behavior occurs for 
intermediate coupling where the broadening $\Gamma$ is comparable to or slightly smaller than the charging 
energy $U$. In this case shown in Fig.\ \ref{fig1}(a), one observes the appearance of {\it new asymmetric}
peaks in the conductance trace. In the absence of disconnected states, there would be only a single 
conductance peak due to the single level of the connected dot. In the presence of the disconnected
states, the occupation of the connected state decreases abruptly whenever a disconnected state is filled up. 
Thus the sharp jump in the conductance is downward (upward) if it occurs on the rising (falling) side 
of the conductance peak of the connected level. This leads to the appearance of new trap-induced 
peaks  in $g$ whose asymmetry arises from the abrupt jumps. Indeed, hints of this
behavior have recently been seen in experiment and were attributed to charging of 
disconnected states \cite{lindemann02}. 

For strongly-connected dots ($U\ll \Gamma$), the jumps in the occupation $n_{\alpha}$ of the connected states
which are associated with charging of disconnected states are typically small compared to one. In this 
case, also the abrupt changes in the conductance are small compared to the conductance itself, leading 
to a characteristic sawtooth-behavior of $g$ as function of chemical potential. This is shown in Fig.\
\ref{fig1}(b). When the connected level is close to half filling, a typical dome shape is observed. 
This is very similar to the behavior seen in recent charge-sensing experiments on a quantum point 
contact monitoring the charge in a disconnected dot \cite{johnson03}.
 
\begin{figure}\centering
\epsfxsize 7cm\epsfbox{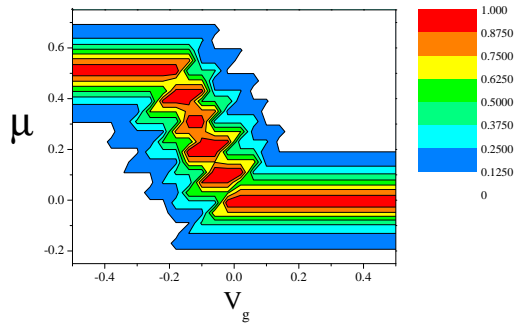} 
\caption{Conductance as function of overall chemical potential $\mu$
and external gate voltage $V_g$ for the disconnected states
(see inset in Fig.\ \ref{fig1}). 
Connected state: $\epsilon=0$;
disconnected states: $e_{m}=-0.55+0.05m$ ($m=1,\ldots,5$),
$U=0.1$ and $t=0.2$. 
} \label{fig2}
\end{figure}

In experiment, the gate voltage does not affect the connected and disconnected levels
in the same way. Using different gates one can even manipulate the levels independently
\cite{johnson03}.
Such experiments correspond to non-vertical trajectories in the $(V_g,\mu)$ plane where $\mu$ is the 
overall chemical potential and $V_g$ is assumed to affect the disconnected states only. 
The resulting intricate pattern of the conductance in the $(V_g,\mu)$ plane is shown in Fig.\ \ref{fig2}.

\begin{figure}\centering
\epsfxsize7cm\epsfbox{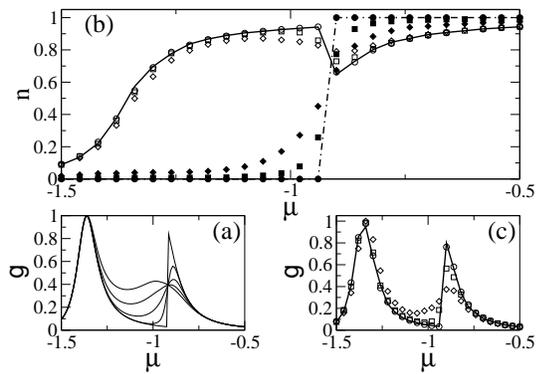} 
\caption{
(a) Temperature dependence of the conductance as function of
chemical potential (connected state: $\varepsilon=-1.3$;
disconnected state: $e_{1}=-1.26$; $U=0.4$; $t=0.3$).
The different curves correspond to $\pi kT=0,0.1,0.2,0.3,0.4$.
(b) Occupation 
of the connected state $n_{\alpha}$ (empty symbols) 
and ``disconnected'' state $n_d$ (filled symbols)
for several values of couplings as the
``disconnected'' state opens up {\it additional} non-current carrying lead.
Lines: results of the exact solution of Eq.\ (\ref{free});
symbols: DMRG calculations for 
different couplings of the ``disconnected'' state 
(circles $t=0$, squares $t=0.1$, diamonds $t=0.2$). 
(c) Conductance of connected state in (b).
} \label{fig3}
\end{figure}

Temperature leads to very interesting behavior of the conductance, even when 
$kT \ll \Gamma$. In the latter regime, the sharp jumps of the conductance 
are broadened by temperature, while the other side of the conductance peaks 
is smooth on the scale $\Gamma$ and thus insensitive to $T$. This results in a very 
asymmetric temperature broadening of the sawtooth-like peaks which was indeed observed 
in experiment \cite{lindemann02}. These considerations can be quantified
by noting that close to the charging of a disconnected state the system can be well 
approximated by a two-level system, with the two levels corresponding to $n_d=0$ and $n_d=1$
(or, more generally, $n_d=N$ and $n_d=N+1$). The occupation of the connected state is then
given by
\begin{eqnarray}
n_{\alpha}={{[{1\over2}+\arctan{2 \mu\over\Gamma}]e^{-\omega_0}+[{1\over 2}+\arctan{2 (\mu-U)\over\Gamma}]
e^{-\omega_1}}\over{e^{-\omega_0}+e^{-\omega_1}}},
\label{temperature}
\end{eqnarray}
where $\omega_{n_d}=\Omega_{n_d}/kT$.
Since the entropy is
governed by the lead states which are unaffected by the change in $n_d$, one expects 
the entropy terms in $\Omega$ to be equal, i.e.,
$T S_1=T S_0$, and thus to cancel out from Eq.\ (\ref{temperature}). 
Expanding $\Omega_1(\mu_{\rm switch}+ \delta \mu)-\Omega_0(\mu_{\rm switch}+ \delta \mu)
\sim kT$ in $\delta\mu$,
one obtains $k T \sim \delta \mu(1-[\arctan(2 (\mu_{\rm switch}-U)/\Gamma)-\arctan
(2 \mu_{\rm switch}/\Gamma)]/\pi$.  Therefore, the abrupt change in $n_{\alpha}$ and hence $g$ is 
smeared by temperature over a range $kT$ in $\mu$.
The resulting $T$ dependence of the conductance is depicted in Fig.\ \ref{fig3}(a). 

A qualitatively similar effect occurs when the disconnected state is broadened, e.g.,
by coupling it to an external reservoir different from the current-carrying
leads. This situation is no longer amenable to an exact solution since  
the occupation of the ``disconnected'' state can now differ from
$0$ or $1$, and we resort to a numerical DMRG method. (We have checked
that the DMRG reproduces our exact solution for vanishing broadening, 
see Fig.\ \ref{fig3}(b).) We find that the main effect of a 
finitely coupled ''disconnected" state is to smear the sudden jump in the occupation of 
the latter, similar to the effect of finite temperature. 
Correspondingly, the effect of the broadening on $g$ 
is very similar to that of temperature,
as can indeed be verified by comparing Figs.\ \ref{fig3}(a) and \ref{fig3}(c).

This observation enables us to qualitatively understand the physics of two connected
states of the same dot with $\Gamma_{1} \gg \Gamma_{2}$.
Indeed, the level occupations are insensitive to whether the levels are coupled to the same or 
different leads. Fig.\ \ref{fig3}(b) then applies and is consistent with very recent works on such
setups employing Hartree-Fock and numerical RG methods \cite{sindel04,konig04}. Evidently,
the way these states are coupled to the leads is crucial for the conductance,  
due to interference effects such as Fano resonances \cite{johnson03,kobayashi04}.

\begin{figure}\centering
\epsfxsize6cm\epsfbox{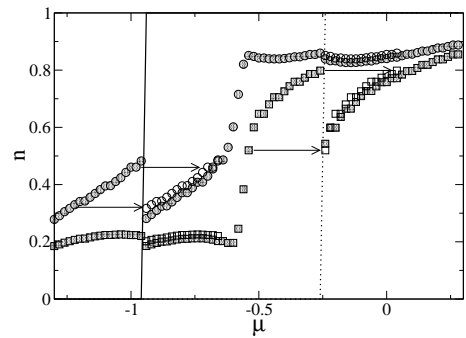}
\caption{
Occupation for two connected states ($t=0.5$) at $\epsilon_1=-1.1$
(gray circles) and $\epsilon_2=-1$ (gray squares) and two disconnected states
$e_1=-1.12$ (line), $e_2=-1.02$ (dotted line) as function of $\mu$
calculated using DMRG for $U=0.3$.
To demonstrate that right after the switch $n_{\alpha}(\mu)=n_{\alpha}(\mu-U)$, we plot
$n_{\alpha}(\mu+U)$ for values of $\mu$ prior to the switch (i.e., $-1.2<\mu<-1$,
and $-0.48<\mu<-0.28$) shown by white symbols.
A clear overlap between the white and gray symbols is seen.
} \label{fig4}
\end{figure}

The insight gained from our exact analysis involving a single connected 
state may be used to make quantitative predictions concerning
{\it several} connected and disconnected states, including charging energies for the connected-dot 
states. Although it is then difficult to calculate the exact switching point for a 
particular disconnected state, one nevertheless predicts that for chemical potentials
$\mu$ (or $V_g$) immediately before or after the switching, $n_{\alpha j}(\mu\pm 0)=n_{\alpha j}(\mu\mp U)$ and therefore
$g(\mu\pm 0)=g(\mu\mp U)$. (Here, $n_{\alpha j}$ denotes the occupation of the $j$th connected state.) 
This behavior is illustrated in Fig.\ \ref{fig4} for two connected and two disconnected 
states, based on a DMRG calculation. This prediction should be very useful for analyzing 
experimental data. For any abrupt jump due to charging of a disconnected
state, the conductance satisfies this relation. By contrast, 
if the jump in the conductance is due e.g.\ to noise no such 
correlation is expected.

In conclusion, it is interesting to speculate that the charge-sensing physics discussed theoretically
in this paper may occur generically in relatively well-coupled chaotic quantum dots. By the nature 
of the Porter-Thomas-distribution of lead-induced level broadenings, there will be a significant number 
of narrow levels in addition to broader levels, in particular for systems with time-reversal symmetry. 
Charge sensing and patterns not unlike Fig.\ \ref{fig1}(b) may thus be important ingredients in explaining the large-scale structure of Coulomb-blockade
sequences observed in such systems \cite{speculation}.

We acknowledge very useful discussions with  I.\ Lerner, C.M.\ Marcus, Y.\ Oreg, and P.\ Silvestrov, 
as well as support from the Israel Academy of Science (YG and RB), SFB 290, the ``Junge Akademie" (FvO) and the BSF (YG). 
One of us (FvO) thanks the Einstein Center (Weizmann) and the Minerva Center  
(Bar-Ilan) for hospitality while part of this work was done.

\end{document}